\documentclass[manyauthors,nocleardouble,COMPASS]{cernphprep}

\usepackage{bm}
\usepackage{color}

\usepackage{enumerate}
\usepackage{amssymb}
\usepackage{amsmath}
\usepackage{amsbsy}
\usepackage{times}
\usepackage{epsfig}
\usepackage{colordvi}
\usepackage{graphicx}
\usepackage{wrapfig,rotating}
\usepackage[numbers, square, comma, sort&compress]{natbib}
\usepackage{multicol}
\usepackage{booktabs}
\usepackage{multirow}
\usepackage{lineno}
\usepackage{wasysym}
\usepackage{graphicx}
\usepackage{mathrsfs}
\usepackage{subfig}
\usepackage{caption} 
\usepackage{tabularx}
\usepackage[percent]{overpic}
\usepackage[a4paper,breaklinks=true]{hyperref} 
\usepackage{url}
\usepackage{tikz}

\RequirePackage[T1]{fontenc}

\usetikzlibrary{plotmarks}

\hypersetup{
     bookmarks=true,         % show bookmarks bar?
     unicode=false,          % non-Latin characters in Acrobat's bookmarks
     pdftoolbar=true,        % show Acrobat's toolbar?
     pdfmenubar=true,        % show Acrobat's menu?
     pdffitwindow=false,     % window fit to page when opened
     pdfstartview={FitH},    % fits the width of the page to the window
     pdfauthor={The COMPASS Collaboration},     % author
     pdfsubject={PWA},   % subject of the document
     pdfnewwindow=true,      % links in new window
     colorlinks=false,       % false: boxed links; true: colored links
     linkcolor=red,          % color of internal links
     citecolor=green,        % color of links to bibliography
     filecolor=magenta,      % color of file links
     urlcolor=cyan           % color of external links
}

\makeatletter

\DeclareSymbolFont{letters}     {OML}{cmm}{m}{it}
\DeclareSymbolFont{symbols}     {OMS}{cmsy}{m}{n}
\DeclareSymbolFont{largesymbols}{OMX}{cmex}{m}{n}
\if00

\newcommand{\MeVcc}{\ensuremath{\textrm{MeV}/c^2}}

\newcommand{\GeV}{\ensuremath{\textrm{GeV}}}
\newcommand{\GeVc}{\ensuremath{\textrm{GeV}/c}}
\newcommand{\GeVcc}{\ensuremath{\textrm{GeV}/c^2}}
\else

\newcommand{\MeVcc}{\ensuremath{\textrm{MeV}}}

\newcommand{\GeV}{\ensuremath{\textrm{GeV}}}
\newcommand{\GeVc}{\ensuremath{\textrm{GeV}}}
\newcommand{\GeVcc}{\ensuremath{\textrm{GeV}}}
\fi

\newcommand{\mm}{\ensuremath{\textrm{mm}}}
\newcommand{\cm}{\ensuremath{\textrm{cm}}}

\newcommand{\ns}{\ensuremath{\textrm{ns}}}

\newcommand{\dd}{\ensuremath{\mathrm{d}}}

\newcommand{\phiGJ}{\ensuremath{\varphi_{\mathrm{GJ}}}}
\newcommand{\thGJ}{\ensuremath{\vartheta_{\mathrm{GJ}}}}
\newcommand{\costhGJ}{\ensuremath{\cos\vartheta_{\mathrm{GJ}}}}
 
\begin{document}

\begin{titlepage}

\PHnumber{2014--204}
\PHdate{12 August 2014}

\title{Odd and Even Partial Waves of $\eta\pi^-$ and
  $\eta'\pi^-$ in $\pi^-p\to\eta^{(\prime)}\pi^-p$ at $191\,\textrm{GeV}/c$}

\Collaboration{The COMPASS Collaboration}
\ShortAuthor{The COMPASS Collaboration}

\begin{abstract}
Exclusive production of $\eta\pi^-$ and $\eta'\pi^-$ has been studied with a
$191\,\GeVc$ $\pi^-$ beam impinging on a hydrogen target at COMPASS (CERN).
Partial-wave analyses reveal different odd/even angular momentum ($L$)
characteristics in the inspected invariant mass range up to $3\,\GeVcc$.  A
striking similarity between the two systems is observed for the $L=2,4,6$
intensities (scaled by kinematical factors) and the relative phases.  The known
resonances $a_2(1320)$ and $a_4(2040)$ are in line with this similarity.  In
contrast, a strong enhancement of $\eta'\pi^-$ over $\eta\pi^-$ is found for the
$L=1,3,5$ waves, which carry non-$q\bar q$ quantum numbers.  The $L=1$ intensity
peaks at $1.7\,\GeVcc$ in $\eta'\pi^-$ and at $1.4\,\GeVcc$ in $\eta\pi^-$, the
corresponding phase motions with respect to $L=2$ are different.
\end{abstract}
\vfill 
\Submitted{(to be submitted to Phys. Lett. B)}

\end{titlepage}

\pagestyle{empty} 
%%%%%%%%%%%%%%%%%%%%%%%%%%%%%%%%%%%%%%%%%%%%%%%%%%%%%%%%%%%%%%%%%%%%%%%%%%%%%%
%
% 2014_auththorlist.tex  
%
%%%%%%%%%%%%%%%%%%%%%%%%%%%%%%%%%%%%%%%%%%%%%%%%%%%%%%%%%%%%%%%%%%%%%%%%%%%%%%

\section*{The COMPASS Collaboration}
\label{app:collab}
\renewcommand\labelenumi{\textsuperscript{\theenumi}~}
\renewcommand\theenumi{\arabic{enumi}}
\begin{flushleft}
C.~Adolph\Irefn{erlangen},
R.~Akhunzyanov\Irefn{dubna}, %phd
M.G.~Alexeev\Irefn{turin_u},
G.D.~Alexeev\Irefn{dubna}, %1
A.~Amoroso\Irefnn{turin_u}{turin_i},
V.~Andrieux\Irefn{saclay},
V.~Anosov\Irefn{dubna}, %2
A.~Austregesilo\Irefnn{cern}{munichtu},
B.~Bade{\l}ek\Irefn{warsawu},
F.~Balestra\Irefnn{turin_u}{turin_i}
% ...
\end{flushleft}

%%%%%%%%%%%%%%%%%%%%%%%%%%%%%%%%%%%%%%%%%%%%%%%%%%%%%%%%%%%%%%%%%%%%%%%%%%%%%%%%%%%%%%%%%%%%%%%%%%%%%%%%%%%%%%%%%%%%%%%
%
% institutes
%
%%%%%%%%%%%%%%%%%%%%%%%%%%%%%%%%%%%%%%%%%%%%%%%%%%%%%%%%%%%%%%%%%%%%%%%%%%%%%%%%%%%%%%%%%%%%%%%%%%%%%%%%%%%%%%%%%%%%%%%

\begin{Authlist}
\item \Idef{bielefeld}{Universit\"at Bielefeld, Fakult\"at f\"ur Physik, 33501 Bielefeld, Germany\Arefs{f}}
\item \Idef{bochum}{Universit\"at Bochum, Institut f\"ur Experimentalphysik, 44780 Bochum, Germany\Arefs{f}\Arefs{ll}}
\item \Idef{bonniskp}{Universit\"at Bonn, Helmholtz-Institut f\"ur  Strahlen- und Kernphysik, 53115 Bonn, Germany\Arefs{f}}
\item \Idef{bonnpi}{Universit\"at Bonn, Physikalisches Institut, 53115 Bonn, Germany\Arefs{f}}
\item \Idef{brno}{Institute of Scientific Instruments, AS CR, 61264 Brno, Czech Republic\Arefs{g}}
\item \Idef{calcutta}{Matrivani Institute of Experimental Research \& Education, Calcutta-700 030, India\Arefs{h}}
\item \Idef{dubna}{Joint Institute for Nuclear Research, 141980 Dubna, Moscow region, Russia\Arefs{i}}
\item \Idef{erlangen}{Universit\"at Erlangen--N\"urnberg, Physikalisches Institut, 91054 Erlangen, Germany\Arefs{f}}
\item \Idef{freiburg}{Universit\"at Freiburg, Physikalisches Institut, 79104 Freiburg, Germany\Arefs{f}\Arefs{ll}}
\item \Idef{cern}{CERN, 1211 Geneva 23, Switzerland}
\item \Idef{liberec}{Technical University in Liberec, 46117 Liberec, Czech Republic\Arefs{g}}
\item \Idef{lisbon}{LIP, 1000-149 Lisbon, Portugal\Arefs{j}}
\item \Idef{mainz}{Universit\"at Mainz, Institut f\"ur Kernphysik, 55099 Mainz, Germany\Arefs{f}}
\item \Idef{miyazaki}{University of Miyazaki, Miyazaki 889-2192, Japan\Arefs{k}}
\item \Idef{moscowlpi}{Lebedev Physical Institute, 119991 Moscow, Russia}
\item \Idef{munichlmu}{Ludwig-Maximilians-Universit\"at M\"unchen, Department f\"ur Physik, 80799 Munich, Germany\Arefs{f}\Arefs{l}}
\item \Idef{munichtu}{Technische Universit\"at M\"unchen, Physik Department, 85748 Garching, Germany\Arefs{f}\Arefs{l}}
\item \Idef{nagoya}{Nagoya University, 464 Nagoya, Japan\Arefs{k}}
\item \Idef{praguecu}{Charles University in Prague, Faculty of Mathematics and Physics, 18000 Prague, Czech Republic\Arefs{g}}
\item \Idef{praguectu}{Czech Technical University in Prague, 16636 Prague, Czech Republic\Arefs{g}}
\item \Idef{protvino}{State Scientific Center Institute for High Energy Physics of National Research Center `Kurchatov Institute', 142281 Protvino, Russia}
\item \Idef{saclay}{CEA IRFU/SPhN Saclay, 91191 Gif-sur-Yvette, France\Arefs{ll}}
\item \Idef{telaviv}{Tel Aviv University, School of Physics and Astronomy, 69978 Tel Aviv, Israel\Arefs{m}}
\item \Idef{triest_u}{University of Trieste, Department of Physics, 34127 Trieste, Italy}
\item \Idef{triest_i}{Trieste Section of INFN, 34127 Trieste, Italy}
\item \Idef{triest_ictp}{Abdus Salam ICTP, 34151 Trieste, Italy}
\item \Idef{turin_u}{University of Turin, Department of Physics, 10125 Turin, Italy}
\item \Idef{turin_p}{University of Eastern Piedmont, 15100 Alessandria, Italy}
\item \Idef{turin_i}{Torino Section of INFN, 10125 Turin, Italy}
\item \Idef{warsaw}{National Centre for Nuclear Research, 00-681 Warsaw, Poland\Arefs{n} }
\item \Idef{warsawu}{University of Warsaw, Faculty of Physics, 00-681 Warsaw, Poland\Arefs{n} }
\item \Idef{warsawtu}{Warsaw University of Technology, Institute of Radioelectronics, 00-665 Warsaw, Poland\Arefs{n} }
\item \Idef{yamagata}{Yamagata University, Yamagata, 992-8510 Japan\Arefs{k} }
\end{Authlist}
%%%%%%%%%%%%%%%%%%%%%%%%%%%%%%%%%%%%%%%%%%%%%%%%%%%%%%%%%%%%%%%%%%%%%%%%%%%%%%%%%%%%%%%%%%%%%%%%%%%%%%%%%%%%%%%%%%%%%%%
%
% Notes
%
%%%%%%%%%%%%%%%%%%%%%%%%%%%%%%%%%%%%%%%%%%%%%%%%%%%%%%%%%%%%%%%%%%%%%%%%%%%%%%%%%%%%%%%%%%%%%%%%%%%%%%%%%%%%%%%%%%%%%%%
\vspace*{-\baselineskip}\renewcommand\theenumi{\alph{enumi}}
\begin{Authlist}
\item \Adef{a}{Also at Instituto Superior T\'ecnico, Universidade de Lisboa, Lisbon, Portugal}
\item \Adef{aa}{Also at Department of Physics, Pusan National University, Busan 609-735, Republic of Korea and at Physics Department, Brookhaven National Laboratory, Upton, NY 11973, U.S.A. }
\item \Adef{bb}{Supported by the DFG Research Training Group Programme 1102  ``Physics at Hadron Accelerators''}
\item \Adef{b}{Also at Chubu University, Kasugai, Aichi, 487-8501 Japan\Arefs{k}}
\item \Adef{c}{Also at KEK, 1-1 Oho, Tsukuba, Ibaraki, 305-0801 Japan}
\item \Adef{c1c}{Present address: Universit\"at Bonn, Helmholtz-Institut f\"ur Strahlen- und Kernphysik, 53115 Bonn, Germany}
\item \Adef{cc}{Also at Moscow Institute of Physics and Technology, Moscow Region, 141700, Russia}
\item \Adef{x}{present address: RWTH Aachen University, III. Physikalisches Institut, 52056 Aachen, Germany}
\item \Adef{f}{Supported by the German Bundesministerium f\"ur Bildung und Forschung}
\item \Adef{g}{Supported by Czech Republic MEYS Grants ME492 and LA242}
\item \Adef{h}{Supported by SAIL (CSR), Govt.\ of India}
\item \Adef{i}{Supported by CERN-RFBR Grants 08-02-91009 and 12-02-91500}
\item \Adef{j}{\raggedright Supported by the Portuguese FCT - Funda\c{c}\~{a}o para a Ci\^{e}ncia e Tecnologia, COMPETE and QREN, Grants CERN/FP/109323/2009, CERN/FP/116376/2010 and CERN/FP/123600/2011}
\item \Adef{k}{Supported by the MEXT and the JSPS under the Grants No.18002006, No.20540299 and No.18540281; Daiko Foundation and Yamada Foundation}
\item \Adef{l}{Supported by the DFG cluster of excellence `Origin and Structure of the Universe' (www.universe-cluster.de)}
\item \Adef{ll}{Supported by EU FP7 (HadronPhysics3, Grant Agreement number 283286)}
\item \Adef{m}{Supported by the Israel Science Foundation, founded by the Israel Academy of Sciences and Humanities}
\item \Adef{n}{Supported by the Polish NCN Grant DEC-2011/01/M/ST2/02350}
\item [{\makebox[2mm][l]{\textsuperscript{*}}}] Deceased
\end{Authlist}

\newpage

The $\eta \pi$ and $\eta' \pi$ mesonic systems are attractive for spectroscopic
studies because any state with odd angular momentum $L$, which coincides with
the total spin $J$, has non-$q\bar q$ (``exotic'') quantum numbers
$J^{PC}=1^{-+},3^{-+},5^{-+},\ldots$ The $1^{-+}$ state has been the principal
case studied so far~\cite{Klempt:2007cp,Brambilla:2014aaa}.

A comparison of $\eta\pi$ and $\eta'\pi$ should illuminate the role of flavour
symmetry.  Since $\eta$ and $\eta'$ are dominantly flavour octet and singlet
states, respectively, different $\mathrm{SU}(3)_\textrm{flavour}$ configurations
are formed by $\eta \pi$ and $\eta' \pi$.  These configurations are linked to
odd or even $L$ by Bose symmetry~\cite{Close:1987aw,Iddir:1988jd,Chung:2002fz}.
Indeed, experimentally the diffractively produced $P$-wave ($L=J=1$) in $\eta'
\pi^-$ was found to be more pronounced than in $\eta
\pi^-$~\cite{Beladidze:1993km}.  A more systematic study of the two systems in
the odd and even partial waves is desirable.

Diffractive production of $\eta \pi^-$ and $\eta' \pi^-$ was studied by previous
experiments with $\pi^-$ beams in the $18\,\GeVc$-$37\,\GeVc$
range~\cite{Beladidze:1993km,Chung:1999we,Ivanov:2001rv,Dorofeev:2001xu}.  Apart
from the well-known resonances $a_2(1320)$ and $a_4(2040)$, resonance features
were observed for the exotic $P$-wave in the $1.4\,\GeVcc-1.7\,\GeVcc$ mass
range.  It has quantum numbers $J^{PG}=1^{--}$, where $G$-parity is used for the
charged system, corresponding to $C=+1$ since the isospin is 1.  Results for
charge-exchange production of $\eta^{(\prime)}\pi^0$ are difficult to relate to
these observations~\cite{Klempt:2007cp}.  Critical discussions of the resonance
character concern a possible dynamical origin of the behaviour of the $L=1$ wave
in these systems~\cite{Donnachie:1998tya,Szczepaniak:2003vg,Klempt:2007cp}.

The present study is performed with a $191\,\GeVc$ $\pi^-$ beam and in the
region $0.1\,(\GeVc)^2 < -t < 1\,(\GeVc)^2$, where $t$ denotes the squared
four-momentum transfer to the proton target.  This is within the range of
Reggeon-exchange processes~\cite{Donnachie:2002en,Shimada:1978sx}, where
diffractive excitation and mid-rapidity (``central'') production coexist.  The
former can induce exclusive resonance production.  The latter will lead to a
system of the leading and the centrally produced mesons with (almost) no
interaction in the final state.

In this Letter, the behaviour of all partial waves with $L=1-6$ in the
$\eta^{(\prime)}\pi^-$ invariant mass range up to $3\,\GeVcc$ is studied.  A
peculiar difference between $\eta \pi^-$ and $\eta' \pi^-$ in the even and
odd-$L$ waves is observed.

The data were collected with the COMPASS apparatus at CERN.  COMPASS is a
two-stage magnetic spectrometer with tracking and calorimetry in both
stages~\cite{Abbon:2007pq,Alexeev:2011}.  A beam of negatively charged hadrons
at $191\, \GeVc$ was impinging on a liquid hydrogen target of $40\,\cm$ length
and $35\,\mm$ diameter.  Using the information from beam particle identifaction
detectors, it was checked that $K^-$ and $\bar p$ admixtures to the 97\% $\pi^-$
beam are insignificant in the final sample analysed here.  Recoiling target
protons were identified by their time of flight and energy loss in a detector
(RPD) which consisted of two cylindrical rings of scintillating counters at
distances of $12\,\cm$ and $78\,\cm$ from the beam axis, covering the polar
angle range above $50^\circ$ as seen from the target centre.  The angular range
between the RPD and the opening angle of the spectrometer of about $\pm
10^\circ$ was covered mostly by a large-area photon and charged-particle veto
detector (SW), thus enriching the data recording with kinematically complete
events~\cite{Schluter:2011}.  The trigger for taking the present data required
coincidence between beam definition counters and the RPD, and no veto from the
SW nor from a small counter telescope for non-interacting beam particles far
downstream ($32\,\textrm{m}$) from the target.  A sample of $4.5\times 10^9$
events was recorded with this trigger in 2008.

For the analysis of the exclusively produced $\pi^-\eta$ and $\pi^-\eta'$
mesonic systems, the $\eta$ was detected by its decay $\eta\to\pi^-\pi^+\pi^0$
($\pi^0 \to\gamma\gamma$), and the $\eta'$ by its decay $\eta'\to\pi^-\pi^+\eta$
($\eta \to\gamma\gamma$).  The preselection for the common final state
$\pi^-\pi^-\pi^+\gamma\gamma$ required
\begin{enumerate}[(a)]
\item three tracks with total charge $-1$ reconstructed in the spectrometer,
\item a vertex, located inside the target volume, with one incoming beam
  particle track and the three outgoing tracks,
\item exactly two ``eligible'' clusters in the electromagnetic calorimeters of
  COMPASS (ECAL1, ECAL2), and
\item the total energy $E_{\rm tot}$ of the outgoing particles within a
  $10\,\GeV$ wide window centred on the $6\,\GeV$ FWHM peak at $191\,\GeV$ in
  the $E_{\rm tot}$ distribution.
\end{enumerate}
Clusters were considered ``eligible'' if they were not associated with a
reconstructed track, if the cluster energy was above 1 \GeV{} and 4 \GeV{} in
ECAL1 and ECAL2, respectively, and if their timing with respect to the beam was
within $\pm 4\,\ns$.

Sharp $\eta$ ($\eta'$) peaks of widths $3\,\MeVcc$-$4\,\MeVcc$ were obtained in
the $\pi^-\pi^+\pi^0$ and $\pi^-\pi^+\eta$ mass spectra after kinematic fitting
of the $\gamma\gamma$ systems within $\pm 20\,\MeVcc$ windows about the
respective $\pi^0$ and $\eta$ masses.  For the present four-body analyses of the
systems $\pi^-\pi^-\pi^+\pi^0$ and $\pi^-\pi^-\pi^+\eta$, broad windows of
$50\,\MeVcc$ width about the $\eta$ and $\eta'$ masses were applied to the
three-body $\pi^-\pi^+\pi^0$ and $\pi^-\pi^+\eta$ systems, respectively.  In
this way, a common treatment of $\eta^{(\prime)}$ and the small number of
non-$\eta^{(\prime)}$ events becomes possible in the subsequent likelihood fit.
No significant deviations from coplanarity (required to hold within $13^\circ$)
are observed for the momentum vectors of beam particle, mesonic system and
recoil proton, which confirms the exclusivity of the reaction.  Details are
found in Refs.~\cite{Schluter:2011b,Schluter:2012}.

In order to account for the acceptance of the spectrometer and the selection
procedure, Monte Carlo simulations~\cite{Alexeev:2011,Geant3} were performed for
four-body phase-space distributions.  The latter were weighted with the
experimental $t$ distributions, approximated by
$\textrm{d}\sigma/\textrm{d}t\propto |t|\exp(-b{}|t|)$ with slope parameter
$b=8.0\,(\GeVc)^{-2}$ and $b=8.45\,(\GeVc)^{-2}$ for $\eta'\pi^-$ and
$\eta\pi^-$, respectively.  The observed weak mass-dependence of the slope
parameter was found not to affect the present results.  The overall acceptances
for $\eta \pi^-$ and $\eta' \pi^-$ in the present kinematic range and decay
channels amounted to $10\,\%$ and $14\,\%$, respectively.  Due to the large
coverage of forward solid angle by the COMPASS spectrometer, the acceptances
vary smoothly over the relevant regions of phase space, see
Ref.~\cite{Schluter:2013}.  A test of the Monte Carlo description was provided
by comparison to a five-charged-track sample where $\eta'$ decays via $\pi^+
\pi^- \eta$ ($\eta\to\pi^+ \pi^- \pi^0$).  The known branching ratio of $\eta$
decay into $\gamma\gamma$ and $\pi^-\pi^+\pi^0$ was
reproduced~\cite{Schluter:2012} leading to a conservative estimate of 8\% for
the uncertainty of the relative acceptance of the two channels discussed here.

\begin{figure}[htbp]
\centering
\subfloat[$m(\eta\pi^-)$]{\includegraphics[width=0.48\textwidth]{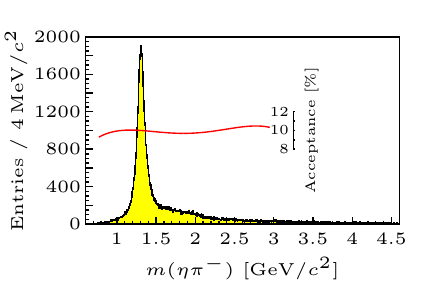}}
\subfloat[$m(\eta\pi^-)$]{\includegraphics[width=0.48\textwidth]{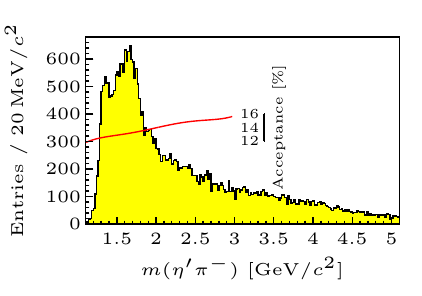}}
\caption{Invariant mass spectra (not acceptance corrected) for (a)
$\eta\pi^-$ and (b) $\eta'\pi^-$.  Acceptances (continuous lines)
refer to the kinematic ranges of the present analysis.}
\label{fig:spectra}
\end{figure}

\begin{figure}[htbp]
\centering
\subfloat[$m(\eta\pi^-)$  vs. $\costhGJ$]{\includegraphics[width=0.48\textwidth]{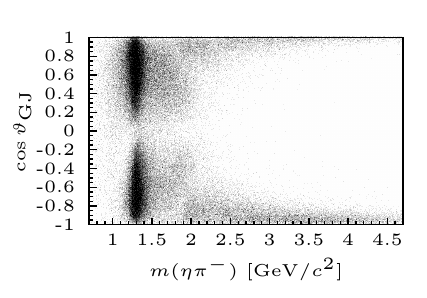}}
\subfloat[$m(\eta'\pi^-)$ vs. $\costhGJ$]{\includegraphics[width=0.48\textwidth]{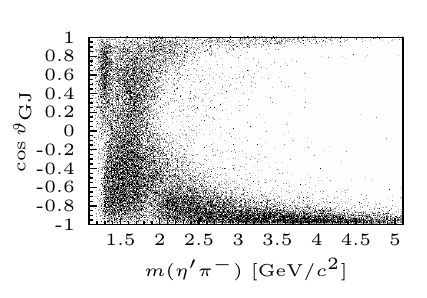}}
\caption{Data (not acceptance corrected) as a function of the
  invariant $\eta\pi^-$ (a) and $\eta'\pi^-$ (b) mass and of the
  cosine of the decay angle in the respective Gottfried-Jackson
  frames where $\cos\thGJ=1$ corresponds $\eta^{(\prime)}$ emission
  in the beam direction.  Two-dimensional acceptances can be found
  in Ref.~\protect\cite{Schluter:2013}.  
\label{fig:massVsCosTh}}
\end{figure}

To visualize the gross features of the two channels, subsamples of events were
selected with tight $\pm 10\,\MeVcc$ windows on the $\eta$ and $\eta'$ masses.
These contain 116\,000 and 39\,000 events, respectively, including $5\%$
background from non-$\eta^{(\prime)}$ events.  These subsamples are shown as
function of the $\eta \pi^-$ and $\eta' \pi^-$ mass in Figs.~\ref{fig:spectra}
(a) and (b), and additonally in the scatter plots Figs.~\ref{fig:massVsCosTh}
(a) and (b) as a function of these invariant masses and of $\costhGJ$, where
$\thGJ$ is the angle between the directions of the $\eta^{(\prime)}$ and the
beam as seen in the centre of mass of the $\eta^{(\prime)}\pi^-$ system (polar
angle in the Gottfried-Jackson frame).  These distributions are integrated over
$|t|$ from $0.1\,(\GeVc)^2$ to $1.0\,(\GeVc)^2$ and over the azimuth $\phiGJ$
(measured with respect to the reaction plane).  The $\phiGJ$ distributions are
observed to follow closely a $\sin^2\phiGJ$ pattern throughout the mass ranges
covered in both channels~\cite{Schluter:2012,Schluter:2013}.

Several salient features of the intensity distributions in
Fig.~\ref{fig:massVsCosTh} are noted before proceeding to the partial-wave
analysis.  In the $\eta \pi^-$ data, the $a_2(1320)$ with its two-hump $D$-wave
angular distribution is prominent, see also Fig.~\ref{fig:spectra} (a).  The
$D$-wave pattern extends to $2\,\GeVcc$ where interference with the $a_4(2040)$
can be discerned.  For higher masses, increasingly narrow forward/backward peaks
are observed.  This feature corresponds to the emergence of a rapidity gap.  In
terms of partial waves it indicates coherent contributions from larger angular
momenta.  Forward/backward asymmetries (only weakly affected by acceptance)
occur for all masses in both channels, which indicates interference of odd and
even partial waves.  In the $\eta' \pi^-$ data, the $a_2(1320)$ is close to the
threshold energy of this channel ($1.1\,\GeV$), and the signal is not dominant,
see also Fig.~\ref{fig:spectra} (b).  A forward/backward asymmetric interference
pattern, indicating coherent $D$- and $P$-wave contributions with mass-dependent
relative phase, governs the $\eta'\pi^-$ mass range up to $2\,\GeVcc$.  In the
$a_4(2040)$ region, well-localised interference is recognised.  As for $\eta
\pi^-$, narrow forward/backward peaking occurs at higher mass, but in this case
the forward/backward asymmetry is visibly larger over the whole mass range of
$\eta' \pi^-$.

The data were subjected to a partial-wave analysis (PWA) using a program
developed at Illinois and
VES~\cite{Ascoli:1970xi,Ryabchikov:private,Alekseev:2009aa}.  Independent fits
were carried out in $40\,\MeVcc$ wide bins of the four-body mass from threshold
up to $3\,\GeVcc$ (so-called mass-independent PWA).  Momentum transfers were
limited to the range given above.

An $\eta^{(\prime)} \pi^-$ partial-wave is characterised by the angular momentum
$L$, the absolute value of the magnetic quantum number $M=|m|$ and the
reflectivity $\epsilon=\pm 1$, which is the eigenvalue of reflection about the
production plane.  Positive (negative) $\epsilon$ is chosen to correspond to
natural (unnatural) spin-parity of the exchanged Reggeon with
$J^P_{\textrm{tr}}=1^-$ or $2^+$ or $3^-$ \dots ($0^-$ or $1^+$ or $2^-$ \dots)
transfer to the beam particle~\cite{Schluter:2012,Chung:1997qd}.  These two
classes are incoherent.

In each mass bin, the differential cross section as a function of four-body
kinematic variables $\tau$ is taken to be proportional to a model intensity
$I(\tau)$ which is expressed in terms of partial-wave amplitudes
$\psi^\epsilon_{LM}(\tau)$,
\begin{equation}
\label{eq:intensity}
I(\tau)=\sum_\epsilon\left|\sum_{L,M}A^\epsilon_{LM}\psi^\epsilon_{LM}(\tau)\right|^2 +
\textrm{non-}\eta^{(\prime)}\textrm{ background}.
\end{equation}
The magnitudes and phases of the complex numbers $A^\epsilon_{LM}$
constitute the free parameters of the fit.  The expected number of
events in a bin is
\begin{equation}
  \bar N \propto \int I(\tau) a(\tau) \dd\tau,
\end{equation}
where $\dd\tau$ is the four-body phase space element and $a(\tau)$
designates the efficiency of detector and selection.  Following the
extended likelihood approach~\cite{Barlow:1990aa,Chung:1997qd}, fits
are carried out maximizing
\begin{equation}
  \label{eq:loglikelihood}
  \ln \mathcal{L} \sim{}  {-} {\bar N} + \sum_{k=1}^n \ln I(\tau_k),
\end{equation}
where the sum runs over all observed events in the mass bin.  In this way, the
acceptance-corrected model intensity is fit to the data.

The partial-wave amplitudes are composed of two parts: a factor $f_\eta$
($f_{\eta'}$) that describes both the Dalitz plot distribution of the successive
$\eta$ ($\eta'$) decay~\cite{Beringer:2012zz} and the experimental peak shape,
and a two-body partial-wave factor that depends on the primary
$\eta^{(\prime)}\pi^-$ decay angles.  In this way, the four-body analysis is
reduced to quasi-two-body.  The partial-wave factor for the two spinless mesons
is expressed by spherical harmonics.  Thus, the full
$\eta(\pi^-\pi^+\pi^0)\pi^-$ partial-wave amplitudes read
\begin{equation}
  \label{eq:amplitude}\begin{split}
  \psi^\epsilon_{LM}(\tau)= &f_\eta(p_{\pi^-},p_{\pi^+},p_{\pi^0})\times
  Y_L^M(\thGJ,0)\\ &\times
  \begin{cases}
    \sin M\phiGJ&\text{for } \epsilon=+1\\ \cos M\phiGJ&\text{for } \epsilon=-1
  \end{cases}\end{split}
\end{equation}
and analoguously for $\eta'(\pi^-\pi^+\eta)\pi^-$.  There are no $M=0$, and
therefore no $L=0$ waves for $\epsilon=+1$.  The fits require a weak $L=M=0$,
$\epsilon=-1$ amplitude which contributes $0.5\%$ ($1.1\%$) to the total
$\eta\pi^-$ ($\eta'\pi^-$) intensity.  This isotropic wave is attributed to
incoherent background containing $\eta^{(\prime)}$, whereas the
non-$\eta^{(\prime)}$ background amplitude in Eq.~\ref{eq:intensity} is
isotropic in four-body phase space.

An independent two-body PWA was carried out not taking into account the decays
of the $\eta^{(\prime)}$, but using tight window cuts ($\pm 10\,\MeVcc)$ on the
$\eta^{(\prime)}$ peak in the respective three-body spectra.  The results were
found to be consistent with the present analysis~\cite{Schluter:2012}.

The above-mentioned azimuthal $\sin^2\phiGJ$ dependence is in agreement with a
strong $M=1$ dominance, as was experienced
earlier~\cite{Beladidze:1993km,Chung:1999we,Ivanov:2001rv,Dorofeev:2001xu}.  No
$M>1$ contributions are needed to fit the data in the present $t$ range, with
the exception of the $\eta \pi^-$ $D$-wave where statistics allows the
extraction of a small $M=2$ contribution.  The final fit model is restricted to
the coherent $L=1-6$, $M=1$ plus $L=2$, $M=2$ partial waves from natural parity
transfer ($\epsilon=+1$) and the incoherent backgrounds introduced above.

Incoherence of partial waves of the same naturality, leading to additional terms
in Eq.~(\ref{eq:intensity}), could arise from contributions with and without
proton helicity flip, or from different $t$-dependences of the amplitudes over
the broad $t$ range.  However, for two pseudoscalars, incoherence or partial
incoherence of any two partial waves with $M=1$ can be accommodated by full
coherence with appropriate choice of phase~\cite{Chung:1999we}.  Comparing PWA
results for $t$ above and below $0.3\,(\GeVc)^2$, no significant variation of
the relative $M=1$ amplitudes with $t$ is observed~\cite{Schluter:2012}.  The
$L=2, M=2$ contribution shows a different $t$-dependence but does not introduce
significant incoherence.

In general, a two-pseudoscalar PWA suffers from discrete
ambiguities~\cite{Martin:1978jn,Sadovsky:1991hm,Chung:1997qd}.  The observed
insignificance of unnatural-parity transfer crucially reduces the ambiguities.
In the case of $\eta \pi^-$, the remaining ambiguities are resolved when the
$M=2$ $D$-wave amplitude is introduced.  For $\eta' \pi^-$, ambiguities occur
when the PWA is extended beyond the dominant $L = 1$, $2$ and $4$ waves.  We
resolve this by requiring continuous behaviour of the dominant partial waves and
of the Barrelet zeros~\cite{Chung:1997qd}.  The acceptable solutions agree
within the statistical uncertainties with the solution selected here, which is
the one with the smallest $L=3$ contribution.

\begin{figure}[htbp]
\vspace{-4em}
\centering
\subfloat[$P$-wave, $L=1$]{     \includegraphics[width=0.48\textwidth]{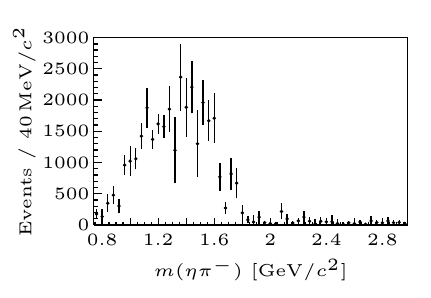}}
\subfloat[$D$-wave, $L=2$]{     \includegraphics[width=0.48\textwidth]{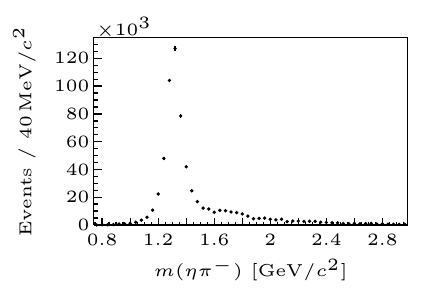}}\\
\subfloat[$F$-wave, $L=3$]{     \includegraphics[width=0.48\textwidth]{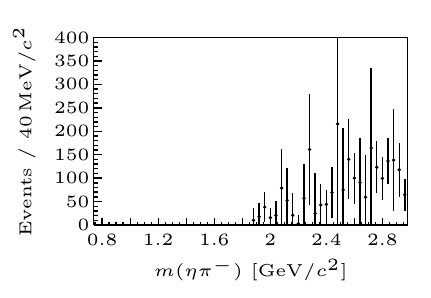}}
\subfloat[$G$-wave, $L=4$]{     \includegraphics[width=0.48\textwidth]{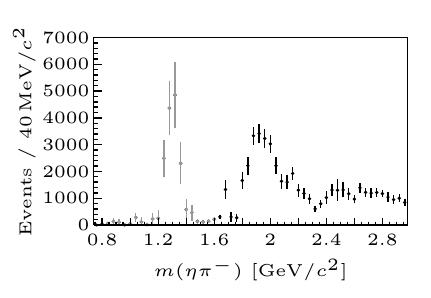}}\\
\subfloat[$H$-wave, $L=5$]{     \includegraphics[width=0.48\textwidth]{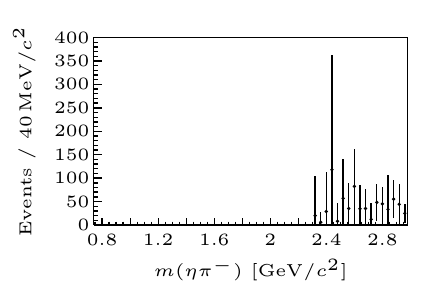}}
\subfloat[$I$-wave, $L=6$]{     \includegraphics[width=0.48\textwidth]{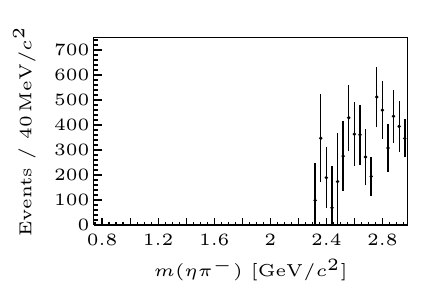}}\\
\subfloat[$D$-wave, $L=2, M=2$]{\includegraphics[width=0.48\textwidth]{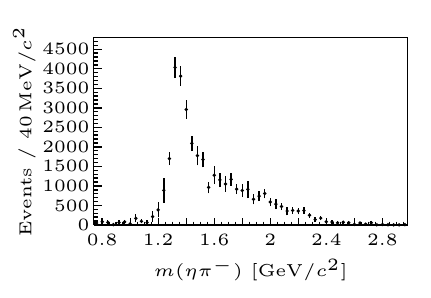}}
\caption{Intensities of the $L=1-6$, $M=1$ and $L=2$, $M=2$ partial
  waves from the partial-wave analysis of the $\eta \pi^-$ data in
  mass bins of $40\,\MeVcc$ width.  The light-colored part of the
  $L=4$ intensity below $1.5\,\GeVcc$ is due to feedthrough from the
  $L=2$ wave.  The error bars correspond to a change of the
  $\log$-likelihood by half a unit and do not include MC
  fluctuations which are on the order of 5\%.}
\label{fig:intensitiesPiEta}
\end{figure}

\begin{figure}[htbp]
  \centering
  \subfloat[$P$-wave, $L=1$]{\includegraphics[width=0.48\textwidth]{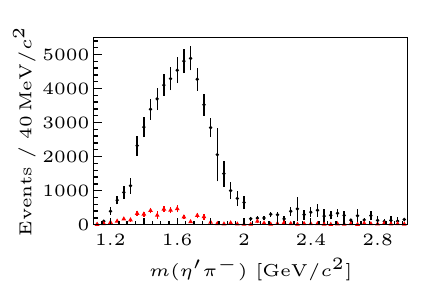}}
  \subfloat[$D$-wave, $L=2$]{\includegraphics[width=0.48\textwidth]{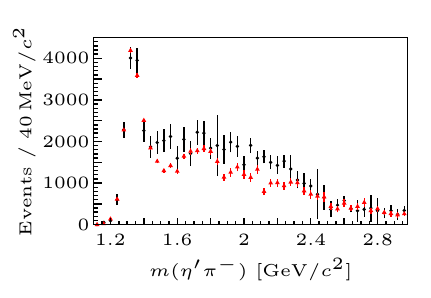}}\\
  \subfloat[$F$-wave, $L=3$]{\includegraphics[width=0.48\textwidth]{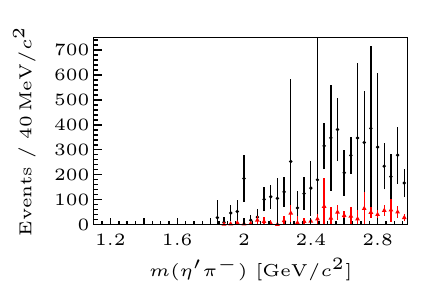}}
  \subfloat[$G$-wave, $L=4$]{\includegraphics[width=0.48\textwidth]{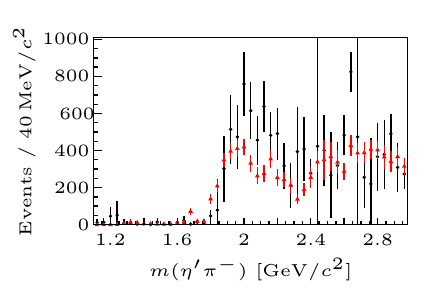}}\\
  \subfloat[$H$-wave, $L=5$]{\includegraphics[width=0.48\textwidth]{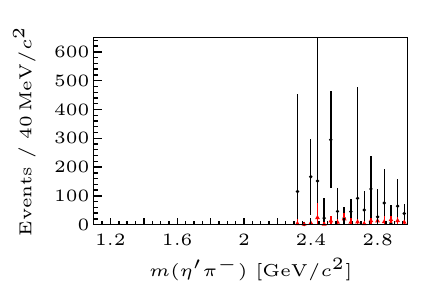}}
  \subfloat[$I$-wave, $L=6$]{\includegraphics[width=0.48\textwidth]{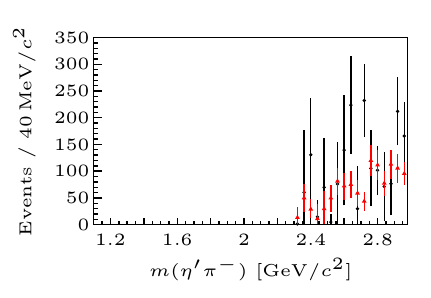}}

  \caption{Intensities of the $L=1-6$, $M=1$ partial waves from the
    partial-wave analysis of the $\eta' \pi^-$ data in mass bins of
    $40\,\MeVcc$ width (circles).  Shown for comparison (triangles)
    are the $\eta \pi^-$ results scaled by the relative kinematical
    factor given in Eq.~\eqref{eq:R}.}
  \label{fig:intensitiesPiEtap}
\end{figure}

\begin{figure}[htbp]
  \centering
  \subfloat[$L=1$]{    \includegraphics[width=0.48\textwidth]{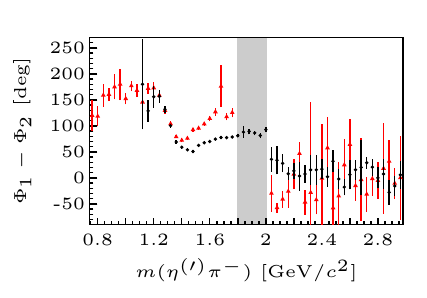}}
  \subfloat[$L=2,M=2$]{\includegraphics[width=0.48\textwidth]{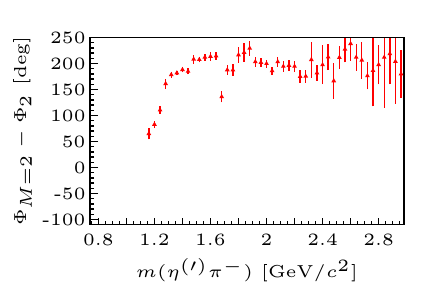}}\\
  \subfloat[$L=3$]{    \includegraphics[width=0.48\textwidth]{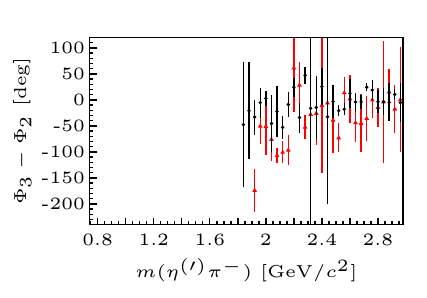}}
  \subfloat[$L=4$]{    \includegraphics[width=0.48\textwidth]{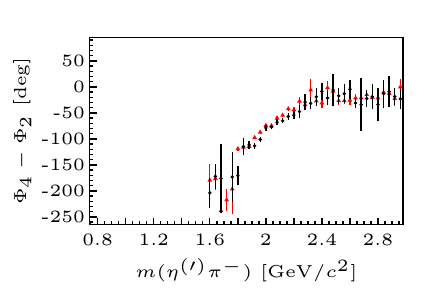}}\\
  \subfloat[$L=5$]{    \includegraphics[width=0.48\textwidth]{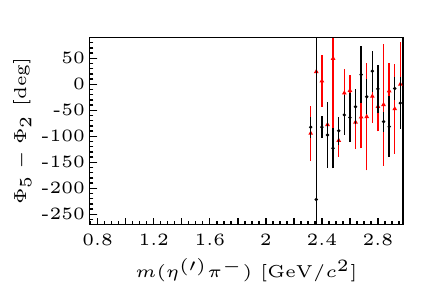}}
  \subfloat[$L=6$]{    \includegraphics[width=0.48\textwidth]{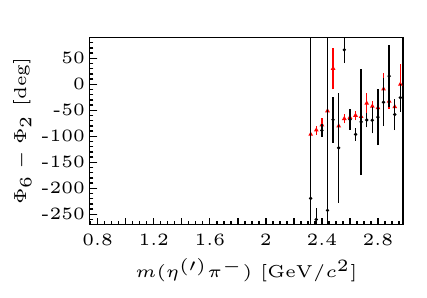}}
  \caption{Phases $\Phi_L$ of the $M=1$ partial waves with angular
    momentum L relative to the $L=2, M=1$ wave of $\eta \pi^-$ (triangles) and
    $\eta' \pi^-$ (circles) systems.  For $\eta \pi^-$, the phase between the
    $P$ and $D$-waves is ill-defined in the region of vanishing $P$-wave
    intensity between 1.8 and $2.05\,\GeVcc$ (shaded).  Panel (b) shows the
    relative $M=2$ versus $M=1$ phase of the $\eta \pi^-$ $D$-wave.}
  \label{fig:phases}
\end{figure}

The results of the PWA are presented as intensities of all included partial
waves in Figs.~\ref{fig:intensitiesPiEta}, \ref{fig:intensitiesPiEtap}, and as
relative phases with respect to the $L=2$, $M=1$ wave in Fig.~\ref{fig:phases}.
The plotted intensities are the acceptance-corrected numbers of events in each
mass bin, as derived from the $|A^\epsilon_{LM}|^2$ of Eq.~\ref{eq:intensity}.
Feedthrough of the order of 3\% from the dominant $a_2(1320)$ signal is observed
in the $L=4$ $\eta \pi^-$ distribution, as shown in light colour in
Fig.~\ref{fig:intensitiesPiEta}.  Relative intensities integrated over mass up
to $3\,\GeVcc$, taking into account the respective $\eta^{(\prime)}$ decay
branchings, are given in Table~\ref{tab:intensities}.  The ratio of the summed
intensities is $I(\eta\pi^-)/I(\eta'\pi^-) = 4.0\pm 0.3$.  This ratio is not
affected by luminosity, its error is estimated from the uncertainty of the
acceptance.  The $\eta\pi^-$ yield is larger for all even-$L$ waves.
Conversely, the odd-$L$ yields are larger in the $\eta'\pi^-$ data.

The $\eta \pi^-$ $P$-wave intensity shows a compact peak of $400\,\MeVcc$ width,
centred at a mass of $1.4\,\GeVcc$.  Beyond $1.8\,\GeVcc$ it disappears.  The
$D$-wave intensity is a factor of twenty larger than the $P$-wave intensity.
These observations resemble those at lower beam
energy~\cite{Chung:1999we,Dorofeev:2001xu}.  A similar $P$-wave peak was
observed in $\bar p n$ annihilation at rest, where it appears with an intensity
comparable to that of the $D$-wave~\cite{Abele:1998gn}.  The present $D$-wave is
characterised by a dominant $a_2(1320)$ peak and a broad shoulder that extends
to higher masses and possibly contains the $a_2(1700)$.  An $M=2$ $D$-wave
intensity is found at the 5\% level.  The $G$-wave shows a peak consistent with
the $a_4(2040)$ and a broad bump centred at about $2.7\,\GeVcc$.  The $F$, $H$
and $I$-waves ($L=3$, 5, 6) adopt each less than 1\% of the intensity in the
present mass range but are significant in the likelihood fit as can be judged
from the uncertainties given in Table~\ref{tab:intensities}.

The $\eta' \pi^-$ $P$ and $D$-wave have comparable intensities.  The former
peaks at $1.65\,\GeVcc$, drops to almost zero at $2\,\GeVcc$ and displays a
broad second maximum around $2.4\,\GeVcc$.  The $D$-wave shows a two-part
structure similar to $\eta\pi^-$ but with relatively larger intensity of the
shoulder.  The $G$-wave distribution shows an $a_4(2040)$ plus bump shape as
observed for $\eta \pi^-$.  In contrast to the $G$ and $I$-waves, the odd $F$
and $H$-waves have an order of magnitude more relative intensity than in the
$\eta\pi^-$ data.  The $F$-wave distribution features a broad peak around
$2.6\,\GeVcc$.

Phase motions in both systems can best be studied with respect to the $D$-wave,
which is present with sufficient intensity in the full mass range.  The rapid
phase rotations caused by the $a_2(1320)$ and $a_4(2040)$ resonances are
discernible.  The $P$ versus $D$-wave phases in both systems are almost the same
from the $\eta'\pi^-$ threshold up to $1.4\,\GeVcc$ where a branching takes
place.  Given the similarity of the $D$-wave intensities after applying a
kinematical factor (see below), it is suggestive to ascribe the different
relative phase motions in the $1.4\,\GeVcc$-$2.0\,\GeVcc$ range to the $P$-wave.
It is noted that the $P$-wave intensities drop dramatically within this region,
almost vanishing at $1.8\,\GeVcc$ in $\eta\pi^-$ and at $2\,\GeVcc$ in
$\eta'\pi^-$.  In contrast, the $G$- versus $D$-phase motions are almost
identical.  All phase differences tend to constant values at high masses, which
is a wave-mechanical condition for narrow angular focussing.

Fits of resonance and background amplitudes to these PWA results (so-called
mass-dependent fits) lead to strongly model-dependent resonance parameters.  If
these fits are restricted to masses below $1.9\, \GeVcc$, comparable to previous
analyses, a simple model incorporating only $P$ and $D$-wave Breit-Wigner
amplitudes and a coherent $D$-wave background yields $\pi_1(1400)$ $\eta \pi^-$
resonance parameters and $\pi_1(1600)$ $\eta' \pi^-$ resonance parameters
consistent with those of
Refs.~\cite{Chung:1999we,Ivanov:2001rv,Dorofeev:2001xu}.  However, the inclusion
of higher masses demands additional model amplitudes, in particular additional
$D$-wave resonances and coherent $P$-wave backgrounds.  The presence of a
coherent background in the $P$-wave is suggested by the PWA results in
Figs.~\ref{fig:intensitiesPiEta}, \ref{fig:intensitiesPiEtap}, \ref{fig:phases}
(a): The vanishing of the intensities around $2.0\,\GeVcc$ is ascribed to
destructive interference within this partial wave, and the relatively slow phase
motion across the $\eta'\pi^-$ $P$-wave peak demands the additional amplitude in
order to dampen the $\pi_1(1600)$ phase rotation.  Fitted $P$-wave resonance
masses in both channels are found to be shifted upwards by typically
$200\,\MeVcc$ when introducing constant-phase model backgrounds as in
Ref.~\cite{Alekseev:2009aa}.  In the present Letter, we refrain from proposing
resonance parameters for the exotic $P$-wave or even the exotic $F$ and
$H$-waves observed here.  The present observations at masses beyond the
$a_2(1320)$ and the $\pi_1$ structures might stimulate extensions of
resonance-production models, as e.g. multi-Regge models~\cite{Shimada:1978sx}.

For the distinct $a_2(1320)$ and $a_4(2040)$ resonances, mass-dependent fits
using a standard relativistic Breit-Wigner parameterisation, which for the $a_2$
includes also the $\rho\pi$ decay in the parameterisation of the total
width~\cite{Beladidze:1993km}, give the following results:
\begin{equation}
  \label{eq:1}
  \begin{split}
  m(a_2) &= 1315\pm 12\,\MeVcc, \quad \Gamma(a_2) = 119\pm 14\,\MeVcc,\\
   m(a_4) &= 1900^{+80}_{-20}\,\MeVcc, \quad
  \Gamma(a_4) = 300^{+80}_{-100}\,\MeVcc, \\
  {B_2} &\equiv \frac{N(a_2\to\eta'\pi^-)}{N(a_2\to\eta\pi)} = (5\pm 2)\%,\\
  \quad B_4 &\equiv \frac{N(a_4\to\eta'\pi^-)}{N(a_4\to\eta\pi)} = (23\pm 7)\%.
\end{split}
\end{equation}
Here, $N$ stands for the integrated Breit-Wigner intensities of the given decay
branches.  The errors given above are dominated by the systematic uncertainty,
which is estimated by comparing fits with and without coherent backgrounds,
$a_2(1700)$ or $\pi_1(1400)$.  The masses and $B_2$ agree with the PDG
values~\cite{Beringer:2012zz}.  The decay branching ratio $B_4$ is extracted
here for the first time.

\begin{table}[htbp]
  \centering

  \caption{Intensities (yields), integrated over the mass range up to
    $3\,\GeVcc$, for the partial waves with $M=1$ (and $M=2$ for $L=2$) relative
    to $L=2,M=1$ in $\eta \pi^-$ (set to 100).  These yields take into account
    the decay branching ratios of $\eta^{(\prime)}$ into
    $\pi^-\pi^+\gamma\gamma$.  Errors are derived from the $\log$-likelihood fit
    and do not include the common uncertainty (8\%) of the acceptance ratio of
    the two channels.  The last column lists $\eta \pi^-$ over $\eta' \pi^-$
    yield ratios derived from the scaled intensities (see text,
    Eq.~\eqref{eq:ratio}).  The first (second) value of $R_L$ corresponds to
    range parameter $r=0\,\textrm{fm}$ ($r=0.4\,\textrm{fm})$.}

   \begin{tabular}{cccc}
         $L$ & yield $(\eta \pi^-)$
         &  yield $(\eta' \pi^-)$& $R_L$ \\
         \hline
         1 &             $5.4\pm 0.3$ &       $12.8\pm 0.4$ & $0.08 - 0.12$ \\
         2 &              100 (fixed) &       $13.0\pm 0.3$ & $0.84 - 1.18$ \\
         2, $M=2$&       $5.4\pm 0.2$ &               & \\
         3 &           $0.39\pm 0.07$ &       $1.14\pm 0.13$ & $0.14 - 0.19$\\
         4 &           $10.0\pm 0.3$  &       $2.57\pm 0.18$ & $0.80 - 0.97$\\
         5 &           $0.12\pm 0.04$ &       $0.28\pm 0.10$ & $0.13 - 0.15$\\
         6 &           $0.87\pm 0.08$ &       $0.36\pm 0.05$ & $0.66 - 0.74$
   \end{tabular}

  \label{tab:intensities}
\end{table}

For a detailed comparison of the results from the mass-independent PWA of both
channels, their different phase spaces and angular-momentum barriers are taken
into account.  For the decay of pointlike particles, transition rates are
expected to be proportional to
\begin{equation}
  \label{eq:scale-factor}
  g(m,L) = q(m)\times q(m)^{2L}
\end{equation}
with break-up momentum $q(m)$~\cite{Peters:1995jv,Abele:1997dz,Bramon:1997va}.
Overlaid on the PWA results for $\eta'\pi^-$ in Fig.~\ref{fig:intensitiesPiEtap}
are those for $\eta \pi^-$, multiplied in each bin by the relative kinematical
factor
\begin{equation}
  \label{eq:R}
  c(m,L) = b\times \frac{g'(m,L)}{g(m,L)},
\end{equation}
where $g^{(\prime)}$ refers to $\eta^{(\prime)} \pi^-$ with break-up momentum
$q^{(\prime)}$, and the factor $b=0.746$ accounts for the decay branchings of
$\eta$ and $\eta'$ into $\pi^-\pi^+\gamma\gamma$~\cite{Beringer:2012zz}.

By integrating the invariant mass spectra of each partial wave, scaled by
$[g^{(\prime)}(m,L)]^{-1}$, from the $\eta'\pi^-$ threshold up to $3\,\GeVcc$,
we obtain scaled yields $I^{(\prime)}_L$ and derive the ratios
\begin{equation}
  \label{eq:ratio}
  R_L=b\times I_L/I'_L.
\end{equation}
As an alternative to the angular-momentum barrier factors $q(m)^{2L}$ of
Eq.~\eqref{eq:scale-factor}, we have also used Blatt-Weisskopf barrier
factors~\cite{VonHippel:1972fg}.  For the range parameter involved there, an
upper limit of $r=0.4\,\textrm{fm}$ was deduced from systematic studies of
tensor meson decays, including the present
channels~\cite{Abele:1997dz,Peters:1995jv}, whereas for $r=0\,\textrm{fm}$
Eq.~\eqref{eq:scale-factor} is recovered.  To demonstrate the sensitivity of
$R_L$ on the barrier model, the range of values corresponding to these upper and
lower limits is given in Table~\ref{tab:intensities}.

The comparison in Fig.~\ref{fig:intensitiesPiEtap} reveals a conspicuous
resemblance of the even-L partial waves of both channels.  This feature remains
if $r=0.4\,\textrm{fm}$, but the values of $R_L$ increase with increasing $r$
(Table~\ref{tab:intensities}).  This similarity is corroborated by the relative
phases as observed in Figs.~\ref{fig:phases} (d) and (f).  The observed
behaviour is expected from a quark-line picture where only the non-strange
components $n \bar n$ ($n=u,d$) of the incoming $\pi^-$ and the outgoing system
are involved.  The similar values of $R_L$ for $L=2,4,6$ suggest that the
respective intermediate states couple to the same flavour content of the
outgoing system.

The quark-line estimate (see Eq.~(3) in~\cite{Abele:1997dz}) for the $a_2(1320)$
decay branching using $r=0.4\,\textrm{fm}$ and the isoscalar mixing angle in the
quark flavour basis, $\phi=39.3^\circ$~\cite{Bramon:1997va}, is $B_2= 3.9\%$ for
our mass value.  This is in reasonable agreement with the present measurement.
An analogous calculation for the $a_4(2040)$ yields $B_4=11.8\%$, which is below
the experimental value.  A larger range parameter $r$ would improve the
agreement.

On the other hand, the odd-$L$ $\eta'\pi^-$ intensities are enhanced by a factor
$5-10$ as compared to $\eta \pi^-$, see Fig.~\ref{fig:intensitiesPiEtap},
Table~\ref{tab:intensities}.  The $P$-wave fits well into the trend observed for
the $F$ and $H$-waves, which also carry exotic quantum numbers.  It is
suggestive to ascribe these observations to the dominant $\mathbf{8}\otimes
\mathbf{8}$ and $\mathbf{1}\otimes \mathbf{8}$ character of the $\eta \pi^-$ and
$\eta' \pi^-$ $SU(3)_\textrm{flavour}$ configurations, respectively.  When the
former couples to an octet intermediate state, Bose symmetry demands even $L$,
whereas the latter may couple to the non-symmetric odd-$L$ configurations.  The
importance of this relation was already pointed out in previous discussions of
the exotic $\pi_1$, where in particular the hybrid ($gq\bar q$) or the lowest
molecular state ($q\bar qq\bar q$) have $\mathbf{1}\otimes \mathbf{8}$
character~\cite{Close:1987aw,Iddir:1988jd,Chung:2002fz}.

A $P$-wave peak, consistent with quoted resonance
parameters~\cite{Beringer:2012zz}, appears in each channel.  In the $\eta'\pi^-$
channel, its relatively large contribution is directly visible in
Fig.~\ref{fig:massVsCosTh}(b).  The forward/backward asymmetry, ascribed to
$L=1,3,5$ amplitudes interfering with the even-$L$ ones, extends to higher
masses, where a transition to rapidity-gap phenomena (central production) is
expected.  In the $\eta\pi^-$ data, the asymmetry is much less pronounced.

In conclusion, two striking features characterise the systematic behaviour of
partial waves presented here:
\begin{enumerate}[(i)]
\item The even partial waves with $L=2,4,6$ show a close similarity between the
  two channels, both in the intensities as function of mass -- after scaling by
  the phase-space and barrier factors -- as well as in their phase behaviour.
\item The odd partial waves with $L=1,3,5$, carrying non-$q\bar q$ quantum
  numbers, are suppressed in $\eta \pi^-$ with respect to $\eta' \pi^-$,
  underlining the importance of flavour symmetry.
\end{enumerate}

\section*{Acknowledgements}

We gratefully acknowledge the support of the CERN management and staff as well
as the skills and efforts of the technicians of the collaborating institutions.
This work was made possible by the financial support of our funding agencies.


\section*{References}
\begin{thebibliography}{10}
\expandafter\ifx\csname url\endcsname\relax
  \def\url#1{\texttt{#1}}\fi
\expandafter\ifx\csname urlprefix\endcsname\relax\def\urlprefix{URL }\fi
\expandafter\ifx\csname href\endcsname\relax
  \def\href#1#2{#2} \def\path#1{#1}\fi

\bibitem{Klempt:2007cp}
E.~Klempt, A.~Zaitsev, {Glueballs, Hybrids, Multiquarks. Experimental facts
  versus QCD inspired concepts}, Phys. Rept. 454 (2007) 1--202.
\newblock \href {http://arxiv.org/abs/hep-ph/0708.4016}
  {\path{arXiv:hep-ph/0708.4016}}, \href
  {http://dx.doi.org/10.1016/j.physrep.2007.07.006}
  {\path{doi:10.1016/j.physrep.2007.07.006}}.

\bibitem{Brambilla:2014aaa}
N.~Brambilla, et~al., {QCD} and strongly coupled gauge theories: challenges and
  perspectives, to be submitted to EPJC\href {http://arxiv.org/abs/1404.3723}
  {\path{arXiv:1404.3723}}.

\bibitem{Close:1987aw}
F.~Close, H.~Lipkin, {New Experimental Evidence for Four Quark Exotics: the
  Serpukhov $\phi \pi$ Resonance and the GAMS $\eta \pi$ Enhancement}, Phys.
  Lett. B196 (1987) 245--250.
\newblock \href {http://dx.doi.org/10.1016/0370-2693(87)90613-7}
  {\path{doi:10.1016/0370-2693(87)90613-7}}.

\bibitem{Iddir:1988jd}
F.~Iddir, et~al., {$q \bar q g$ hybrid and $q q \bar q \bar q$ diquonium
  interpretation of the {GAMS} $1^{-+}$ resonance}, Phys.Lett. B205 (1988)
  564--568.
\newblock \href {http://dx.doi.org/10.1016/0370-2693(88)90999-9}
  {\path{doi:10.1016/0370-2693(88)90999-9}}.

\bibitem{Chung:2002fz}
S.~U. Chung, E.~Klempt, J.~G. K{\"o}rner, {SU(3) classification of p-wave $\eta
  \pi$ and $\eta' \pi$ systems}, Eur. Phys. J. A15 (2002) 539--542.
\newblock \href {http://arxiv.org/abs/hep-ph/0211100}
  {\path{arXiv:hep-ph/0211100}}, \href
  {http://dx.doi.org/10.1140/epja/i2002-10058-0}
  {\path{doi:10.1140/epja/i2002-10058-0}}.

\bibitem{Beladidze:1993km}
G.~Beladidze, et~al., {Study of $\pi^- N \to \eta \pi^- N$ and $\pi^- N \to
  \eta'\pi^- N$ reactions at $37\,\textrm{GeV}/c$}, Phys. Lett. B313 (1993)
  276--282.
\newblock \href {http://dx.doi.org/10.1016/0370-2693(93)91224-B}
  {\path{doi:10.1016/0370-2693(93)91224-B}}.

\bibitem{Chung:1999we}
S.~Chung, et~al., {Evidence for exotic $J^{PC} = 1^{-+}$ meson production in
  the reaction $\pi^- p \to \eta \pi^- p$ at $18\,\textrm{GeV}/c$}, Phys. Rev.
  D60 (1999) 092001.
\newblock \href {http://arxiv.org/abs/hep-ex/9902003}
  {\path{arXiv:hep-ex/9902003}}, \href
  {http://dx.doi.org/10.1103/PhysRevD.60.092001}
  {\path{doi:10.1103/PhysRevD.60.092001}}.

\bibitem{Ivanov:2001rv}
E.~I. Ivanov, et~al., {Observation of exotic meson production in the reaction
  $\pi^- p \to \eta' \pi^- p$ at $18\,\textrm{GeV}/c$}, Phys. Rev. Lett. 86
  (2001) 3977--3980.
\newblock \href {http://arxiv.org/abs/hep-ex/0101058}
  {\path{arXiv:hep-ex/0101058}}, \href
  {http://dx.doi.org/10.1103/PhysRevLett.86.3977}
  {\path{doi:10.1103/PhysRevLett.86.3977}}.

\bibitem{Dorofeev:2001xu}
V.~Dorofeev, et~al., {The $J^{PC} = 1^{-+}$ hunting season at VES}, AIP Conf.
  Proc. 619 (2002) 143--154.
\newblock \href {http://arxiv.org/abs/hep-ex/0110075}
  {\path{arXiv:hep-ex/0110075}}, \href {http://dx.doi.org/10.1063/1.1482444}
  {\path{doi:10.1063/1.1482444}}.

\bibitem{Donnachie:1998tya}
A.~Donnachie, P.~Page, {Interpretation of experimental $J^{PC}$ exotic
  signals}, Phys. Rev. D58 (1998) 114012.
\newblock \href {http://arxiv.org/abs/hep-ph/9808225}
  {\path{arXiv:hep-ph/9808225}}, \href
  {http://dx.doi.org/10.1103/PhysRevD.58.114012}
  {\path{doi:10.1103/PhysRevD.58.114012}}.

\bibitem{Szczepaniak:2003vg}
A.~Szczepaniak, M.~Swat, A.~Dzierba, S.~Teige, {$\eta \pi$ and $\eta' \pi$
  Spectra and Interpretation of Possible Exotic $J^{PC} = 1^{-+}$ Mesons},
  Phys. Rev. Lett. 91 (2003) 092002.
\newblock \href {http://arxiv.org/abs/hep-ph/0304095}
  {\path{arXiv:hep-ph/0304095}}, \href
  {http://dx.doi.org/10.1103/PhysRevLett.91.092002}
  {\path{doi:10.1103/PhysRevLett.91.092002}}.

\bibitem{Donnachie:2002en}
S.~Donnachie, H.~G. Dosch, O.~Nachtmann, P.~Landshoff, {Pomeron Physics and
  QCD}, Camb. Monogr. Part. Phys. Nucl. Phys. Cosmol., Cambridge University
  Press, 2002.

\bibitem{Shimada:1978sx}
T.~Shimada, A.~Martin, A.~Irving, Double regge exchange phenomenology, Nucl.
  Phys. B142 (1978) 344.
\newblock \href {http://dx.doi.org/10.1016/0550-3213(78)90209-2}
  {\path{doi:10.1016/0550-3213(78)90209-2}}.

\bibitem{Abbon:2007pq}
P.~Abbon, et~al., {The COMPASS Experiment at CERN}, Nucl. Inst. Meth. A577
  (2007) 455--518.
\newblock \href {http://arxiv.org/abs/hep-ex/0703049}
  {\path{arXiv:hep-ex/0703049}}, \href
  {http://dx.doi.org/10.1016/j.nima.2007.03.026}
  {\path{doi:10.1016/j.nima.2007.03.026}}.

\bibitem{Alexeev:2011}
M.~Alekseev, et~al., {The COMPASS 2008 Spectrometer}, to be submitted to Nucl.\
  Instr.\ and Meth.\ A (2014).

\bibitem{Schluter:2011}
T.~Schl{\"u}ter, et~al., {Large-Area Sandwich Veto Detector with WLS Fibre
  Readout for Hadron Spectroscopy at COMPASS}, Nucl. Inst. and Meth. A654
  (2011) 219.
\newblock \href {http://arxiv.org/abs/1108.4587} {\path{arXiv:1108.4587}},
  \href {http://dx.doi.org/10.1016/j.nima.2011.05.069}
  {\path{doi:10.1016/j.nima.2011.05.069}}.

\bibitem{Schluter:2011b}
T.~Schl{\"u}ter, \href{http://www.slac.stanford.edu/econf/C110613}{{The exotic
  $\eta'\pi^-$ Wave in $190\,\textrm{GeV}$ $\pi^-p\to\pi^-\eta'p$ at COMPASS}},
  in: B.~Grube, S.~Paul, N.~Brambilla (Eds.), Proceedings of the XIV
  International Conference on Hadron Spectroscopy, eConf C110613, 2011.
\newblock \href {http://arxiv.org/abs/1108.6191} {\path{arXiv:1108.6191}}.
\newline\urlprefix\url{http://www.slac.stanford.edu/econf/C110613}

\bibitem{Schluter:2012}
T.~Schl{\"u}ter,
  \href{{http://wwwcompass.cern.ch/compass/publications/theses/2012\%5Fphd\%5F%
schlueter.pdf}}{The $\pi^-\eta$ and $\pi^-\eta'$ systems in exclusive
  $190\,\textrm{GeV}$ $\pi^-p$ reactions at {COMPASS (CERN)}}, Ph.D. thesis,
  Ludwig-Maximilians-Universit{\"a}t, M{\"u}nchen (2012).
\newline\urlprefix\url{{http://wwwcompass.cern.ch/compass/publications/theses/%
2012\%5Fphd\%5Fschlueter.pdf}}

\bibitem{Geant3}
CERN, GEANT -- Detector Description and Simulation Tool (October 1994).

\bibitem{Schluter:2013}
T.~Schl{\"u}ter, {Odd and Even Partial Waves of $\eta\pi^-$ and $\eta'\pi^-$ in
  $191\,\textrm{GeV}$ $\pi^-p$}, PoS(Hadron 2013) 085.
\newblock \href {http://arxiv.org/abs/1401.4067} {\path{arXiv:1401.4067}}.

\bibitem{Ascoli:1970xi}
G.~Ascoli, et~al., {Partial Wave Analysis of the $3 \pi$ Decay of the $A_2$},
  Phys. Rev. Lett. 25 (1970) 962.
\newblock \href {http://dx.doi.org/10.1103/PhysRevLett.25.962}
  {\path{doi:10.1103/PhysRevLett.25.962}}.

\bibitem{Ryabchikov:private}
I.~Karchaev, D.~Ryabchikov, private communication.

\bibitem{Alekseev:2009aa}
M.~Alekseev, et~al., {Observation of a $J^{PC} = 1^{-+}$ exotic resonance in
  diffractive dissociation of $190\,\textrm{GeV}/c$ $\pi^-$ into $\pi^- \pi^-
  \pi^+$}, Phys. Rev. Lett. 104 (2010) 241803.
\newblock \href {http://arxiv.org/abs/0910.5842} {\path{arXiv:0910.5842}},
  \href {http://dx.doi.org/10.1103/PhysRevLett.104.241803}
  {\path{doi:10.1103/PhysRevLett.104.241803}}.

\bibitem{Chung:1997qd}
S.~U. Chung, {Techniques of amplitude analysis for two pseudoscalar systems},
  Phys. Rev. D56 (1997) 7299--7316.
\newblock \href {http://dx.doi.org/10.1103/PhysRevD.56.7299}
  {\path{doi:10.1103/PhysRevD.56.7299}}.

\bibitem{Barlow:1990aa}
R.~Barlow, Extended maximum likelihood, Nucl. Inst. Meth. A297~(3) (1990) 496
  -- 506.
\newblock \href {http://dx.doi.org/10.1016/0168-9002(90)91334-8}
  {\path{doi:10.1016/0168-9002(90)91334-8}}.

\bibitem{Beringer:2012zz}
J.~Beringer, et~al., {Review of Particle Physics}, Phys.Rev. D86 (2012) 010001.
\newblock \href {http://dx.doi.org/10.1103/PhysRevD.86.010001}
  {\path{doi:10.1103/PhysRevD.86.010001}}.

\bibitem{Martin:1978jn}
A.~Martin, et~al., {A Study of Isospin 1 Meson States Using
  $10\,\textrm{GeV}/c$ $K^- K^0$ Production Data}, Phys. Lett. B74 (1978) 417.
\newblock \href {http://dx.doi.org/10.1016/0370-2693(78)90693-7}
  {\path{doi:10.1016/0370-2693(78)90693-7}}.

\bibitem{Sadovsky:1991hm}
S.~Sadovsky, On the ambiguities in the partial wave analysis of $\pi^-p\to
  \eta\pi^0 n$ reaction, Tech. rep., IHEP 91-75, IHEP Protvino (1991).

\bibitem{Abele:1998gn}
A.~Abele, et~al., {Exotic $\eta\pi$ state in $\bar p d$ annihilation at rest
  into $\pi^- \pi^0 \eta p_{\textrm{spectator}}$}, Phys. Lett. B423 (1998)
  175--184.
\newblock \href {http://dx.doi.org/10.1016/S0370-2693(98)00123-3}
  {\path{doi:10.1016/S0370-2693(98)00123-3}}.

\bibitem{Peters:1995jv}
K.~Peters, E.~Klempt, {The suppression of $s \bar s$ pair creation from tensor
  meson decays}, Phys.Lett. B352 (1995) 467--471.
\newblock \href {http://dx.doi.org/10.1016/0370-2693(95)00457-V}
  {\path{doi:10.1016/0370-2693(95)00457-V}}.

\bibitem{Abele:1997dz}
A.~Abele, et~al., {Study of the $\pi^0 \pi^0 \eta' final$ state in $\bar p p$
  annihilation at rest}, Phys. Lett. B404 (1997) 179--186.
\newblock \href {http://dx.doi.org/10.1016/S0370-2693(97)00526-1}
  {\path{doi:10.1016/S0370-2693(97)00526-1}}.

\bibitem{Bramon:1997va}
A.~Bramon, R.~Escribano, M.~Scadron, {The $\eta - \eta'$ mixing angle
  revisited}, Eur. Phys. J. C7 (1999) 271--278.
\newblock \href {http://arxiv.org/abs/hep-ph/9711229}
  {\path{arXiv:hep-ph/9711229}}, \href
  {http://dx.doi.org/10.1007/s100529801009} {\path{doi:10.1007/s100529801009}}.

\bibitem{VonHippel:1972fg}
F.~von Hippel, C.~Quigg, {Centrifugal-barrier effects in resonance partial
  decay widths, shapes, and production amplitudes}, Phys. Rev. D5 (1972)
  624--638.
\newblock \href {http://dx.doi.org/10.1103/PhysRevD.5.624}
  {\path{doi:10.1103/PhysRevD.5.624}}.

\end{thebibliography}
\end{document}